\documentclass[12pt,a4paper]{article}
\pdfoutput=1

\usepackage{amsmath,amssymb}
\usepackage[bookmarks=true]{hyperref}
\usepackage[nosort]{cite}
\usepackage[bulletsep]{collect}
\def\gfxon{\usepackage[final]{graphicx}}

\gfxon

\makeatletter
\let\old@startsection=\@startsection
\renewcommand{\@startsection}[6]{\old@startsection{#1}{#2}{#3}{#4}{#5}{#6\mathversion{bold}}}
\makeatother

\newcommand{\dpou}[1]{\partial^{#1}}
\newcommand{\dpod}[1]{\partial_{#1}}

\makeatletter \@addtoreset{equation}{section} \makeatother

\makeatletter
\let\old@makecaption=\@makecaption
\def\@makecaption{\small\old@makecaption}
\makeatother

\makeatletter
\newcommand{\ellSN}{\mathop{\operator@font sn}\nolimits}
\newcommand{\ellCN}{\mathop{\operator@font cn}\nolimits}
\newcommand{\ellDN}{\mathop{\operator@font dn}\nolimits}
\newcommand{\ellAM}{\mathop{\operator@font am}\nolimits}
\newcommand{\ellK}{\mathop{\smash{\operator@font K}\vphantom{a}}\nolimits}
\newcommand{\ellE}{\mathop{\smash{\operator@font E}\vphantom{a}}\nolimits}
\makeatother

\ifx\genfrac\sdflkaj

\else

\fi
\newcommand{\sfrac}[2]{{\textstyle\frac{#1}{#2}}}
\newcommand{\half}{\sfrac{1}{2}}

\newcommand{\xvec}{\textbf{x}}

\newcommand{\nln}{\nonumber\\}
\newcommand{\nl}[1][0pt]{\nonumber\\[#1]&\hspace{-4\arraycolsep}&\mathord{}}

\newcommand{\earel}[1]{\mathrel{}&\hspace{-2\arraycolsep}#1\hspace{-2\arraycolsep}&\mathrel{}}
\newcommand{\eq}{\earel{=}}
\newcommand{\beq}{\begin{equation}}
\newcommand{\eeq}{\end{equation}}

\def\[{\begin{equation}}
\def\]{\end{equation}}
\def\<{\begin{eqnarray}}
\def\>{\end{eqnarray}}

\makeatletter
\def\mr@ignsp#1 {\ifx\:#1\@empty\else #1\expandafter\mr@ignsp\fi}%
\newcommand{\multiref}[1]{\begingroup
\xdef\mr@no@sparg{\expandafter\mr@ignsp#1 \: }%
\def\mr@comma{}%
\@for\mr@refs:=\mr@no@sparg\do{\mr@comma\def\mr@comma{,}\ref{\mr@refs}}%
\endgroup}
\makeatother

\newcommand{\hypref}[2]{\ifx\href\asklfhas #2\else\href{#1}{#2}\fi}

\newcommand{\secref}[1]{Sec.~\multiref{#1}}

\newcommand{\figref}[1]{Fig.~\multiref{#1}}
\renewcommand{\eqref}[1]{(\multiref{#1})}
\newcommand{\lagr}{\mathcal{L}}

\ifx\href\asklfhas\newcommand{\href}[2]{#2}\fi
\newcommand{\arxivno}[1]{\href{http://arxiv.org/abs/#1}{#1}}

\begin{document}

\begin{flushright}\footnotesize
\texttt{ArXiv:\arxivno{1007.1428}}\\
\texttt{UMN-TH-2905/10}\\
\vspace{0.5cm}
\end{flushright}

\begin{center}%
{\Large\textbf{\mathversion{bold}
On the Null Energy Condition and Causality in Lifshitz Holography}
\par}

\vspace{1cm}%

\textsc{Carlos Hoyos$^a$ and Peter Koroteev$^b$}

\vspace{10mm}

\textit{$^a$Department of Physics, University of Washington Seattle \\%
WA 98195-1560, USA}

\vspace{6mm}

\textit{$^b$University of Minnesota, School of Physics and Astronomy\\%
MN 55455, USA}

\vspace{5mm}

\texttt{$^*$choyos@phys.washington.edu, $^\dagger$koroteev@physics.umn.edu}

\vspace{7mm}

\thispagestyle{empty}

\par\vspace{1cm}

\vfill

\textbf{Abstract}\vspace{5mm}

\begin{minipage}{12.7cm}

We use a WKB approximation to establish a relation between the wavefront velocity in a strongly coupled theory and the local speed of light in a holographic dual, with our main focus put on systems with Lifshitz scaling with dynamical exponent $z$. We then use Einstein equations to relate the behavior of the local speed of light in the bulk with the null energy condition (NEC) for bulk matter, and we show that it is violated for Lifshitz backgrounds with $z<1$. We study signal propagation in the gravity dual and show that violations of the NEC are incompatible with causality in the strongly coupled theory, ruling out as holographic models Lifshitz backgrounds with $z<1$. We argue that causality violations in $z<1$ theories will show up in correlators as superluminal modes and confirm this for a particular example with $z=1/2$. Finally, as an application, we use $z<1$ solutions to uncover regions of the parameter space of curvature squared corrections to gravity where the NEC can be violated.
\end{minipage}

\end{center}
\vspace*{\fill}

\newpage

\section{Introduction}\label{sec:Intro}

The AdS/CFT correspondence \cite{Maldacena:1997re, Gubser:1998bc} has been extensively used as a tool to extract properties of strongly coupled systems. In its usual formulation the strongly coupled dynamics of a large-$N$ gauge theory is extracted from a gravitational theory in a higher dimensional space. Recent developments are directed towards the development of similar techniques for strongly coupled critical points appearing in condensed matter systems, a nice introduction to the subject can be found in refs.~\cite{Hartnoll:2009sz,Herzog:2009xv}. Although critical points show some kind of scale invariance, Lorentz symmetry is usually broken so the time coordinate can scale differently to the space coordinates
\[\label{eq:DinamicalScaling}
t\to \lambda^z t\,, \quad \textbf{x}\to \lambda\, \textbf{x}\,.
\]
Here $z$ is known as the dynamical critical exponent. In spatially anisotropic systems where rotational invariance is broken one or more spatial coordinates can also scale differently to the rest. Systems with dynamical scaling \eqref{eq:DinamicalScaling} have been studied for a long time in condensed matter theory \cite{Lifshitz:1941aa}. 

In holographic models with dynamical scaling the starting point is to construct a geometry whose isometries and local symmetries map to the global symmetries of the critical point of interest, most importantly to the scaling properties. Examples where this was 
first done are refs.~\cite{Son:2008ye, Kachru:2008yh, Adams:2008wt}. 
So far theories with known holographic duals are still deformations of large-$N$ gauge theories, but for properties that do not depend strongly on the microscopic details of the theory one wants to study, these holographic constructions could produce good qualitative results that would be very difficult to obtain using other methods. 

In a relativistic theory  the dispersion relation for massless particles is such that the frequency is proportional to the momentum $\omega = c k$. Theories with broken Lorentz invariance $z\neq 1$ \eqref{eq:DinamicalScaling} can have different dispersion relations at small frequencies, so in general an ultraviolet completion would be needed in order to make them compatible with causality. Consider, for instance, a mean field description of massless scalar fluctuations around a critical point with dynamical exponent $z$. The effective Lagrangian would be
\[\label{eq:lifshitzlagr}
\lagr=(\dpod{t}\phi)^2-c^2\ell^{2(z-1)} \phi(-\dpod{\xvec}^2)^z\phi\,,
\]
where $\ell$ has units of length and $c$ is the speed of light. The dispersion relation reads
\[
\omega^2 = {c^2\over \ell^2}(\ell k)^{2 z}\,.
\]
It follows that the phase velocity is
\[\label{eq:phasevel}
v_{\rm ph}={\omega \over k} = c (\ell k)^{z-1} \,.
\]
We could think of $1/\ell$ as a cutoff in $k$, at $k=1/\ell$ fluctuations reach the speed of light and for larger values the theory should have a relativistic description. Although strictly speaking the phase velocity does not have to be smaller than the speed of light, in the limit $\omega\to \infty$ (so $k\to \infty$) the phase velocity becomes the wavefront velocity, that should not be larger than the speed of light $c$. The limit where relativistic effects can be neglected corresponds to $c\to\infty$\footnote{Strictly speaking we rescale the time direction so the speed of light seems to grow.} and $\ell\to 0$ while keeping $\kappa=c\ell^{z-1}$ fixed.

In the discussion above we have made the implicit assumption that $z$ is an integer, otherwise the term with spatial derivatives would not be local. Fractional values of $z$ may appear in more complicated situations.
We should also remark that if $z<1$ the previous argument fails, the phase velocity diverges when $k\to 0$ and does not when $k\to \infty$. Clearly a theory with dynamical exponent $z<1$ cannot be an infrared fixed point, although it is not ruled out that such a theory describes an intermediate range of scales in some system, as for instance this could be the case for  fully developed turbulence \cite{1991RSPSA.434....9K}.

The metric proposed for a holographic description of a critical point with dynamical exponent $z$ is \cite{Kachru:2008yh,Koroteev:2007yp}
\[\label{Metric:KLKLM}
ds^2={L^2\over r^2} \left(-{\kappa^2 dt^2\over r^{2(z-1)}}+ dr^2+d\textbf{x}^2\right)\,.
\]
Here $L$ is the radius of curvature of the metric and there is a conformal boundary at $r=0$. One can apply a holographic recipe adapted to these spaces to compute two-point functions in the dual theory \cite{Kachru:2008yh, Gordeli:2009vh}. The results for scalar operators agree with expectations from scale invariance, in particular they depend on the combination $\omega/(\kappa\, k^z)$.

Now let us make the following observation: the local speed of light at a fixed value of the radial coordinate has a dependence on $1/r$ that is the same as the phase velocity in \eqref{eq:phasevel} with $k$, namely
\[
c(r)=c\,\ell^{(z-1)} r^{-(z-1)}\,.
\]
This observation fits well with our intuition of how holography should work, the radial coordinate is associated with different scales, and radial slices to the field theory at those scales. We will make this argument more rigorous and actually show that the local speed of light is always related to the phase velocity at small wavelengths even in situations where there is no exact scale invariance.

The next point we will address is what is the difference between $z>1$ and $z<1$ from the gravitational perspective. Although both cases have different singular behavior, it is not {\em a priori} clear why one choice would be `better' than the other. We will show that $z<1$ backgrounds are incompatible with causality in the holographic dual, and that this will be true in general for any geometry with a local speed of light decreasing towards the boundary. We will show that violations of causality in this sense are produced by matter that violates the null energy condition (NEC). 

We also study two-point functions of scalar operators computed following the holographic recipe. We write the equation of motion for scalar fields as a Sch\"roedinger equation to argue that the qualitative behavior in $z<1$ and $z>1$ geometries is completely different. For $z<1$ and any finite value of $k$, the potential is confining so the holographic two-point function should show a discrete set of poles. When $k\to 0$ the poles merge forming a branch cut. We give an explicit expression for $z=1/2$ that confirms our arguments.

The current paper is organized as follows. In \secref{Sec:LocalSpeed} we study the relation between local speed of light and phase velocity in the limit where the phase velocity becomes equal to the wavefront velocity. We then argue about causality in Lifshitz backgrounds for $z>1$ and $z<1$ and find very different structure in those cases. \secref{Sec:ScalarCorr} is devoted to the calculation of two-point functions for scalar operators using holographic techniques. We find poles in the propagators for different critical exponents and analyze spectra of scalar perturbations. Having established the importance of the NEC for the holography we can use it to constrain some known theories. In \secref{Sec:NECHighDer} we study higher derivative gravity theories which are also used in holographic constructions and determine a region where Lifshitz solutions with $z<1$ exist and hence violations of the NEC are possible. In \secref{Sec:Conclusions} we present our conclusions and discuss some open questions.

\section{Null energy condition and causality}\label{Sec:LocalSpeed}

We will use a simple setup with a scalar operator in a $d+1$ dimensional field theory with a $D=d+2$ gravity dual. 
The holographic description of the vacuum is given by a background metric and fields in the gravity dual. Assuming rotational invariance in the spatial directions, we can write the metric as
\[\label{eq:gaussmetric}
ds^2=du^2+e^{2 A(u)}(-e^{2 B(u)}dt^2+d\textbf{x}^2)\,,
\]
where $t,\,\textbf{x}$ correspond to the time and space coordinates in the field theory and $u$ is the holographic radial coordinate. In this coordinate system the boundary is located at $u\to\infty$.

We can now introduce a source for the operator for a finite interval of time and let the system relax. If the perturbation is small enough, so linear response theory holds, the final state will simply consist of some scalar modes propagating through the vacuum. One can then expand in plane waves to study the dispersion relation of the scalar modes. The phase velocity is given by the ratio of frequency and momentum $v_{\rm ph}=\omega/k$. In the $\omega\to \infty$ limit, the phase velocity becomes the wavefront velocity $v_{\rm wf}$, that for a relativistic theory should be smaller or equal to the speed of light $v_{\rm wf}\leq 1$. In a non-relativistic theory the velocity can take any value, but it should remain finite at large frequencies if the theory is to have a relativistic completion, as has been discussed in the introduction.

\begin{figure}
\begin{center}
\includegraphics[width=10cm]{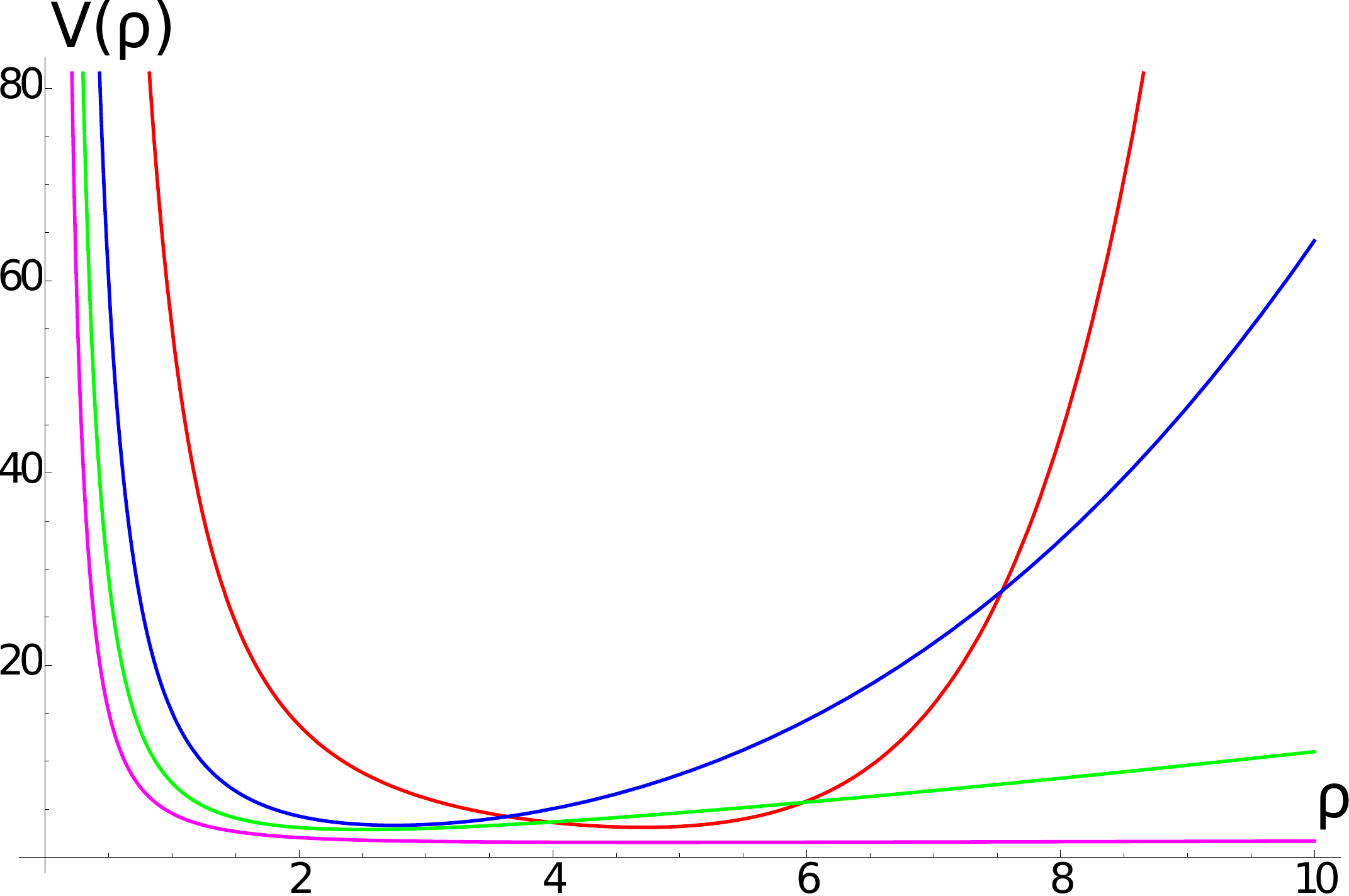}
\caption{Quantum mechanical potentials for $z=0.2, 0.4, 0.6, 0.8$ as functions of the radial coordinate $\rho$. The plots are ordered by their slopes at large $\rho$, the steepest curve corresponds to $z=0.8$. Spatial momentum $k$ is set to unity together with the bulk curvature scale $L$.}\label{fig:potentials}
\end{center}
\end{figure}

In the holographic description the states created by a scalar operator correspond to classical normalizable solutions of a dual scalar field. We have just to consider the quadratic part of the action
\[\label{eq:ScalarFieldAction}
 S = -\int\,d^{d+2} x \sqrt{-g}\left(\dpod{M}\Phi\, \dpou{M}\Phi+m^2\Phi^2\right)\,,
\]
where the mass $m$ depends on the scaling dimensions of the operator. Using a plane wave ansatz $\Phi(t,\textbf{x},u)=e^{-i\omega t+i \textbf{k} \textbf{x}}\phi(u)$ the equations of motion read
\[\label{eq:KGequation}
\phi''+((d+1) A'+B') \phi'+e^{-2 A-2 B}\omega^2\phi - e^{-2 A}k^2 \phi-m^2\phi=0\,.
\]
It is useful to rewrite this equation as a Schr\"odinger equation. First we define a new coordinate 
\[
\rho'(u)=-e^{-A(u)-B(u)}\,.
\]
Then we can rewrite \eqref{eq:KGequation} as follows (derivatives with respect to $\rho$ will be denoted by dots)
\[\label{eq:EOMScalarPlanar}
\ddot{\phi} +d \dot{A} \dot{\phi}+\omega^2\phi -k^2 e^{2 B} \phi-m^2 e^{2 A+2 B}\phi=0\,.
\]
Now define $\phi=\sigma \psi$ with
\[
{\dot{\sigma}\over \sigma}=-{d\over 2} \dot{A}\,.
\]
The equation \eqref{eq:EOMScalarPlanar} then becomes
\[
-\ddot{\psi}+V(\rho)\psi=\omega^2\psi\,,
\]
which is a Schr\"odinger equation with $\omega^2$ playing the role of the energy. The potential is\footnote{Similar analysis was performed in refs.~\cite{Dubovsky:2001fj, Koroteev:2009xd} where braneworlds with broken Lorentz invariance in the bulk were considered.}
\beq\label{eq:schpot}
V(\rho)=k^2 e^{2B}+m^2 e^{2A+2B}+{d^2\over 4}\dot{A}^2+{d\over 2}\ddot{A}\,.
\eeq
In order to make the construction more explicit let us consider the Lifshitz metric. Then in the notations of \eqref{eq:gaussmetric} we have
\beq
A(u)={u\over L}\,,\quad B(u)=(z-1){u\over L}\,,
\eeq
and the variable $\rho=Le^{-zu/L}/z$, such that $\rho \to 0$ is the boundary limit and $\rho\to \infty$ is the horizon limit, this is equivalent to doing the change of coordinates $\rho= r^z/z$ in \eqref{Metric:KLKLM}. 
We rescale the time coordinate so $\omega^2\to \omega^2/\kappa^2$. Then the potential \eqref{eq:schpot} reads
\beq\label{eq:Potentials}
V(\rho)=\frac{d^2+2 d z+4 m^2L^2}{4 \rho^2 z^2}+k^2\left(\frac{\rho z}{L}\right)^{\frac{2}{z}-2}\,.
\eeq
If we use the relation between the mass and the scaling dimension of the dual operator  $m^2L^2=\Delta (\Delta-d-z)$, the potential becomes
\beq\label{eq:schreq}
V(\rho)=\frac{1}{\rho^2}\left[{1\over z^2}\left(\Delta-{d+z\over 2}\right)^2-{1\over 4}+{k^2\over z^2}(z \rho)^{2/z} \right]\,.
\eeq
The potentials \eqref{eq:Potentials} for $d=2$ and $m^2=0$ are plotted in \figref{fig:potentials} for several values of $z$.
The first term in \eqref{eq:Potentials} is singular near the boundary, while the behavior of the second term depends on the value of $z$.
By simple inspection of the potential we can see that solutions have a different qualitative behavior depending on whether $z>1$ or $z<1$.  If $z>1$, the potential decays as $\rho\to\infty$, so the solutions become plane waves $\psi\sim e^{\pm i \omega \rho}$ and the spectrum of normalizable solutions is continuous. For $z<1$ and $k\neq 0$ the potential has a barrier at large values of $\rho$. There is a minimum at
\[\label{eq:rmin}
\rho_{min} = {L\over z }\left(\frac{(1-z)(d^2+2 d z+4 m^2L^2)}{4 z (k L)^2}\right)^{z\over 2}\,.
\]
Notice that for $z=1$ and $k^2>\omega^2$, solutions are either exponentially growing or decreasing. Only the latter can belong to the physical spectrum of the theory, that should be normalizable (or delta-normalizable) over the entire range of the $\rho$ coordinate. For smaller values of $z$ ($z<1$) and $k\neq 0$ the barrier is steeper, so normalizable solutions will be suppressed as well. At $\rho=0$ some negative values of the mass allow two possible normalizable solutions (when the coefficient of the $1/\rho^2$ term is between $3/4$ and $-1/4$), but in holographic applications each choice corresponds to a different boundary theory \cite{Balasubramanian:1998de,Klebanov:1999tb}. Imposing normalizability at both ends leads to a discrete spectrum.

\paragraph{Phase velocity from the WKB approximation.}

Consider the limit of large momentum $k\to \infty$ and large frequency $\omega\to \infty$. The term proportional to $k^2$ in the potential \eqref{eq:schpot} dominates, except very close to the boundary $\rho=0$. In this limit we can use the WKB approximation to the solutions. Notice that $e^B$ is the function that gives the local speed of light at a fixed value of the radial coordinate. If we assume that it decreases, then for large values of $\rho$  the potential will be smaller than the energy $\omega^2> V(\rho)$ and the WKB solution will be oscillatory
\[\label{eq:wkb1}
\psi(\rho) = {c_1\over\sqrt{\pi}\left(\omega^2-V(\rho) \right)^{1/4}} e^{i\int^{\rho} d\tilde{\rho}\, \sqrt{\omega^2-V(\tilde{\rho})} }+ {c_2\over\sqrt{\pi}\left(\omega^2-V(\rho) \right)^{1/4}} e^{-i\int^{\rho} d\tilde{\rho}\, \sqrt{\omega^2-V(\tilde{\rho})}}\,.
\]
However, for values of $\rho$ very close to the boundary $V(\rho)\sim m^2 e^{2A}>\omega^2$. This means that there is a turning point $\rho_0$ where $V(\rho_0)=\omega^2$. In the interval where $V(\rho)>\omega^2$ and the $k^2$ term in the potential still dominates, the WKB solution will be an exponential
\[\label{eq:wkb2}
\psi(\rho) = {d_1\over\sqrt{\pi}\left(V(\rho)-\omega^2 \right)^{1/4}} e^{\int^{\rho} d\tilde{\rho}\, \sqrt{V(\tilde{\rho})-\omega^2} }+ {d_2\over\sqrt{\pi}\left(V(\rho)-\omega^2 \right)^{1/4}} e^{-\int^{\rho} d\tilde{\rho}\, \sqrt{V(\tilde{\rho})-\omega^2} }\,.
\]
As $\rho\to 0$ the potential is too steep $V(\rho)\simeq \gamma/ \rho^2$ for the WKB approximation to be valid (see for instance \cite{Landau:1991} for a treatment of both the WKB approximation and the $1/\rho^2$ potential). As long as $\gamma>-1/4$ (as we will assume it is the case), one can simply do a Frobenius expansion at $\rho=0$, pick the normalizable solution and then match to the WKB solution \eqref{eq:wkb2}. The solution at large values of $\rho$ \eqref{eq:wkb1} can be fixed by matching both WKB solutions at the turning point $\rho_0$ with Airy functions.

We are not really interested in finding the solution, but rather in the value of the turning point in the limit $\omega\to \infty$ we are taking. The condition  $V(\rho_0)=\omega^2$ gives us
\[\label{eq:match}
v_{wf}\simeq v_{ph}={\omega\over k} \simeq  e^{B(\rho_0)}\,.
\]
We see that in this limit the wavefront velocity becomes equal to $e^{B(\rho_0)}$, which is the local speed of light at the turning point. Therefore plane wave states created by a scalar operator in the field theory have wavefront velocities that are equal to the local speed of light in the holographic dual. This confirms the observation we made in the introduction in relation to the local speed of light in Lifshitz geometries.

\paragraph{Growing versus decreasing speed of light.}

In the previous paragraph we assumed that the local speed of light on a radial slice decreases when the radial position is moved towards larger distances from the boundary, and we have assumed $m^2\geq 0$ as well. In principle this argument can be generalized to cover more cases, the potential will be slightly more complicated and whether the solution is oscillating or an exponential on different radial intervals can vary. However, the matching condition \eqref{eq:match} that gives the relation between phase velocity and local speed of light would be the same in the $\omega\to \infty$, $k\to \infty$ limit.

It is interesting to study the difference between geometries with a local speed of light that grows towards the boundary and geometries where it decreases. The Lifshitz metric gives examples for both, depending on whether $z>1$ (growing) or $z<1$ (decreasing). We have already commented in the introduction that field theories with dynamical exponent $z>1$ or $z<1$ have qualitatively different dispersion relations, in the former the phase velocity grows with momentum, while in the latter it decreases. We have just shown how this is reflected in the local speed of light in the geometry. 

A consequence of this difference in dispersion relations is that the boundary structure of $z>1$ and $z<1$ geometries is quite different. We shall recall here the Lifshitz metric
\[
ds^2={L^2\over r^2} \left( -{\kappa^2 dt^2\over r^{2(z-1)}}+dr^2+d\textbf{x}^2\right)\,.
\]
For $z>1$ we do the change of variables $R=r^z$, $t\to t/z$, $\textbf{x}\to \textbf{x}/z$
\beq
ds^2={L^2 \over z^2R^2}\left(-\kappa^2 dt^2+dR^2+R^{2-2/z} d\textbf{x}^2\right)\,. 
\eeq
The conformal boundary is at $R\to 0$. Notice that for $z>1$ it is timelike and one-dimensional, or in other words it is along the time direction. For a general value of $z$ it is also singular. For instance, in the particular case of $z=2$ there is a conical singularity. When $z\to \infty$ the geometry is $AdS_2\times \mathbb{ R}^d$, and the boundary is regular. 

For $z<1$, we can write the metric in the original coordinates as
\beq
ds^2={L^2\over r^2}\left(dr^2+d\textbf{x}^2-r^{2-2 z}\kappa^2 dt^2\right)\,.
\eeq
In this case the conformal boundary $r\to 0$ is a $d$-dimensional surface. In general the boundary is also singular, for instance when $z=1/2$ the singularity is conical. For $z=1$, when the geometry is $AdS_{d+2}$ the boundary is regular and it also includes time.

Notice also that for $z>1$ the slope of null geodesics in the $(t,r)$ plane is such that they are orthogonal to the boundary, while for $z<1$ they are tangent, so the singularity is null instead of timelike. Indeed, for $z<1$ the slope becomes infinite at the boundary but it is reached in a finite time (for $z>0$), the equation for a null geodesic is 
\beq
{dt\over dr}=- \frac{r^{z-1}}{\kappa}\,,\quad  t(r_0)=0\,,
\eeq
which leads to
\[
t(r)= {r_0^z-r^z\over z\kappa} \,.
\]
The presence of a singularity at the boundary, or equivalently at large scales in the dual theory, suggests that an ultraviolet completion is necessary\footnote{This does not mean that the bulk theory is not well defined in the presence of singularities. Although classically there is a singularity, the motion of quantum test particles could be well behaved \cite{Horowitz:1995gi}.}. A possibility is to cut the geometry at a small value of $r$, $r=\ell\ll L$. For Lifshitz backgrounds with $z>1$ we could identify $\ell$ with the cutoff in the non-relativistic theory and $\kappa/\ell^{z-1}$ with the speed of light of the relativistic completion. In some cases \cite{Gubser:2009qt,Goldstein:2009cv,Hartnoll:2009ns,Azeyanagi:2009pr} the geometry is only Lifshitz close to the horizon and becomes an $AdS$ space close to the boundary, so the ultraviolet completion has already been included in the description. A different approach is to try to define the holographic theory without changing the asymptotic boundary behavior or introducing a cutoff, as in  ref. \cite{Horava:2009vy} where the notion of conformal boundary is extended to anisotropic Weyl transformations. A well defined prescription requires to constrain the allowed fluctuations in the bulk, in some cases this implies constraining the sources of the dual theory \cite{Guica:2010sw}. A formal treatment clarifying these issues is certainly desirable. For our purpose it will be enough to introduce a cutoff, since the results we will obtain are independent on how the ultraviolet theory is defined.

A natural question is whether a consistent holographic description requires further conditions. For instance, we have commented in the introduction that Lifshitz theories with $z<1$ would have a bad infrared behavior. However, the boundary analysis of the holographic dual does not seem to be so helpful here. Both $z>1$ and $z<1$ geometries have a singular boundary, and we would introduce a cutoff to get rid of it. In the following we will use causality arguments to show that even with a boundary cutoff the $z<1$ geometries are not good holographic duals.

\paragraph{Causality from shock waves.}

\begin{figure}[htb!]
\begin{center}
\includegraphics[width=7cm]{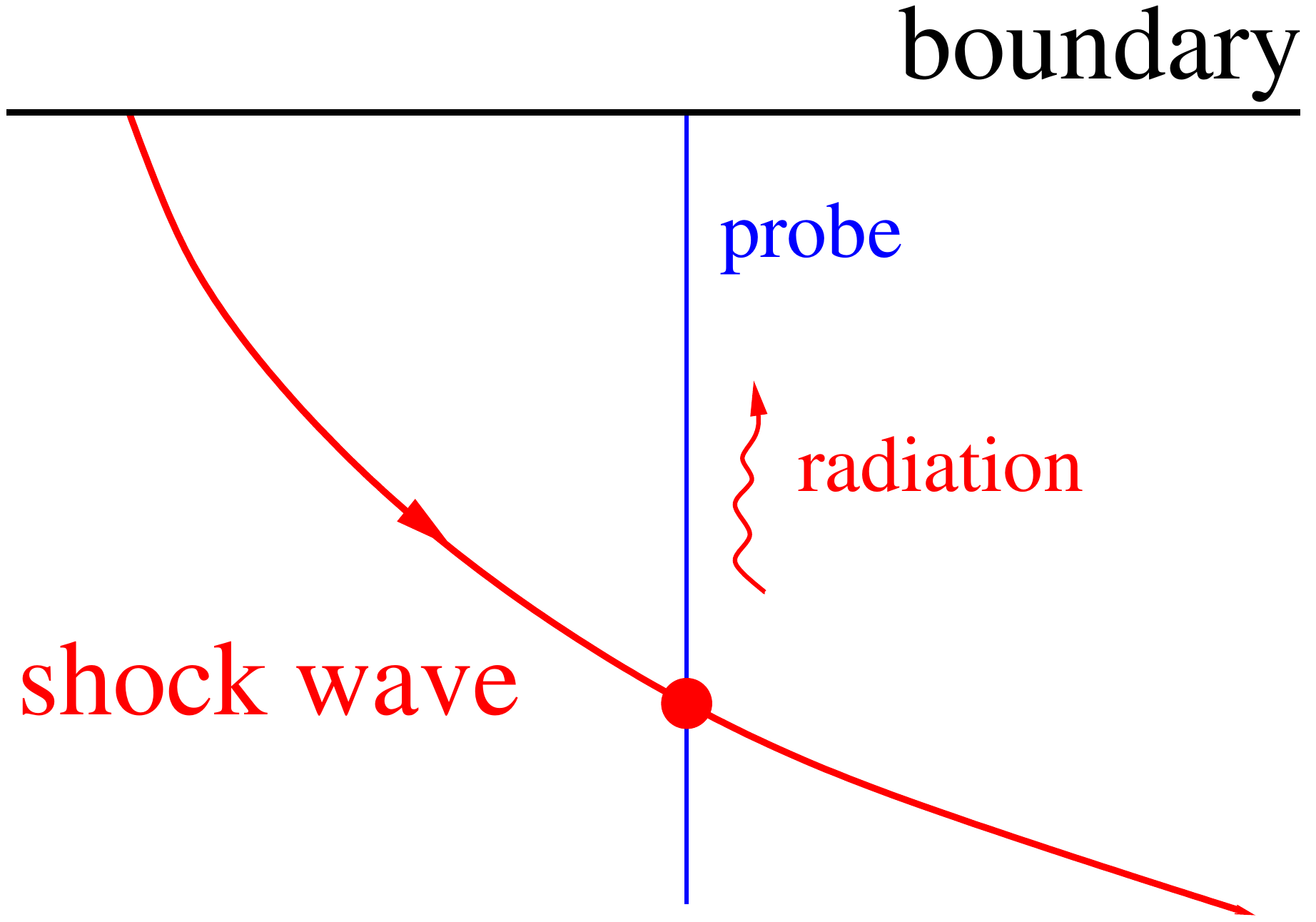} 
\caption{\label{fig:shock}  A source at the boundary produces a shock wave propagating through the  bulk. When the shock wave crosses a static probe a pulse of radiation is sent to the boundary. In the holographic dual the front of radiation produced by the source is determined by the the time and position of the emitted bulk radiation when it reaches the boundary.}
\end{center}
\end{figure}

We will now perform an analysis in the spirit of the conformal colliders thought experiments of Hofman and Maldacena \cite{Hofman:2008ar}. We will introduce a source in the field theory localized in time and in one of the spatial directions $x$. This will produce a planar perturbation propagating along $x$. In the holographic description the source is a boundary condition that will produce a shock wave-like perturbation propagating along null geodesics in both the radial and the spatial direction. This corresponds to both the propagation of the wave and its spreading to larger wavelengths. The shock wave will be a source of radiation of gravitational fields that will then propagate along the radial direction to the boundary, producing a front of radiation that can be interpreted as the front of the perturbation in the dual theory. A similar picture will arise if one uses a probe dragging string, as has been done to study jet physics in holographic duals  \cite{Chesler:2008uy,Chesler:2008wd,Hatta:2008tx}. An alternative way of measuring the effect of the shock wave is to have static probes consisting of strings extended in the radial direction. When the shock wave crosses a probe, it will produce a perturbation that will propagate towards the boundary along radial null geodesics producing a signal that can be interpreted as the position of the front. We have illustrated this picture in \figref{fig:shock}.

We can find the null geodesics\footnote{Recall that in \cite{Koroteev:2009xd} null geodesics were used for qualitative studying localization of field fluctuations in the bulk.} by solving the variational problem with Lagrangian
\beq
{\cal L}=-{\kappa^2 \dot{t}^2 \over r^{2 z}}+{\dot{x}^2+\dot{r}^2\over r^2}\,,
\eeq  
and imposing the constraint ${\cal L}=0$. Since ${\cal L}$ does not depend on $t$ or $x$, we can introduce two integration constants that solve $\dot{t}$ and $\dot{x}$ in terms of $r$
\beq
\dot{t}= {E r^{2 z}\over \kappa^2}\,,\quad \dot{x}=P r^2\,.
\eeq
If we now solve the constraint
\beq
\dot{r}=r^2\sqrt{{ E^2 r^{2(z-1)}\over \kappa^2}-P^2}\,.
\eeq
Then the geodesic equations in terms of $r$ read
\beq\label{eq:geod}
{dt\over dr}= {E r^{2(z-1)}\over \kappa^2\sqrt{{ E^2 r^{2(z-1)}\over \kappa^2}-P^2}}\,,\quad  {dx\over dr}= {P \over \sqrt{{ E^2 r^{2(z-1)}\over \kappa^2}-P^2}}\,.
\eeq
Let us assume that the initial point for the geodesic describing the shock wave is at $r=\ell$, $t=0$, $x=0$. In order to have a sensible solution we need the momentum to be timelike $p^2\equiv {E^2\ell^{2(z-1)}\over \kappa^2}-P^2 >0$. We see now a clear different qualitative behavior between $z>1$ and $z<1$. 

For $z>1$ the argument inside the square root in \eqref{eq:geod} is always positive, so the geodesic extends to all values of $r$. The asymptotic behavior at large $r$ is
\beq\label{eq:tandr}
t\simeq {r^z\over z\kappa} \to \infty\,, \quad x \simeq {\kappa P\over (2-z) E} r^{2-z} + x_0\,.
\eeq 
For the special value $z=2$, $x$ grows logarithmically with $r$. We see that for $z>2$ signals at the boundary can reach only a finite distance that depends on the ratio $P/E< \ell^{(z-1)}/\kappa$. 

For $z<1$ the argument inside the square root in \eqref{eq:geod} becomes negative at a finite value of the radial coordinate, when $r_0^{1-z}= E/(\kappa |P|)$, or using $\kappa=c \ell^{z-1}$ and assuming $P>0$,
\beq\label{eq:r0}
\left(\frac{r_0}{\ell}\right)^{1-z}=\frac{E}{c P}.
\eeq
Although the slope diverges, the values of both $t$ and $x$ approach a finite value, so the null geodesic is tangent to the slice of constant $r$ at that point, and it will bounce back towards the boundary. In the above formula $r_0$ represents the right turning point of the quantum mechanical potential \eqref{eq:schreq}.

The local speed of light at $r=\ell$ is $c$, that is also the speed of light of the dual theory, we can compare this value with the average velocity of the shock wave. Using \eqref{eq:geod} and \eqref{eq:r0}, the space interval that the shock wave has traveled when it comes back to the boundary is
\beq
\Delta x=2\frac{c P}{E}\int_\ell^{r_0} dr \frac{(r/\ell)^{1-z}}{\sqrt{1-\frac{c^2 P^2}{E^2}\left(\frac{r}{\ell}\right)^{2(1-z)}}} = 2\int_\ell^{r_0} dr \frac{(r/r_0)^{1-z}}{\sqrt{1-\left(\frac{r}{r_0}\right)^{2(1-z)}}}.
\eeq 
Similarly, the time interval is
\beq
c \Delta t=2\int_\ell^{r_0} dr \frac{(r/\ell)^{z-1}}{\sqrt{1-\frac{c^2 P^2}{E^2}\left(\frac{r}{\ell}\right)^{2(1-z)}}} = 2\frac{c P}{E}\int_\ell^{r_0} dr \frac{(r/r_0)^{z-1}}{\sqrt{1-\left(\frac{r}{r_0}\right)^{2(1-z)}}}\,.
\eeq 
Let us now assume that $r_0\gg \ell$, so the turning point is located far apart from the boundary. Then we can take the limit $\ell\to 0$ in the above integrals, and after changing the variables $\rho=r/r_0$ we get
\beq
\Delta x =   2 r_0 \int_0^1 d \rho \frac{\rho^{1-z}}{\sqrt{1-\rho^{2(1-z)}}} =2 r_0\frac{\sqrt{\pi} \Gamma\left(\frac{2-z}{2(1-z)} \right)}{ \Gamma\left(\frac{1}{2(1-z)} \right)}\,,
\eeq
\beq
c \Delta t  =  2\frac{c P}{ E} r_0 \int_0^1 d \rho \frac{\rho^{z-1}}{\sqrt{1-\rho^{2(1-z)}}}=2\frac{c P}{ E} r_0\frac{\sqrt{\pi} \Gamma\left(\frac{z}{2(1-z)} \right)}{ (2 z-1)\Gamma\left(\frac{2 z-1}{2(1-z)} \right)} \,.
\eeq
Then the shock wave velocity is
\beq
\frac{v_S}{c} = \frac{\Delta x}{c \Delta t} = z\frac{E}{c P}\,.
\eeq
Since $r_0\gg\ell$, due to \eqref{eq:r0} the latter ratio is bigger than unity, thus the shock wave travels faster than light signals at the boundary $v_S>c$. One can show that the front of radiation will coincide with the shock wave at the boundary, so its average velocity is also larger than the boundary speed of light. As was argued in ref.~\cite{Amado:2008hw} for asymptotically AdS spaces, whenever there is a null geodesic returning to the boundary there are singularities in the correlation functions that correspond to poles in the Fourier transform. Similar arguments should apply in this case, but the poles will correspond to superluminal modes, we will show this explicitly in \secref{Sec:ScalarCorr}. Superluminal propagation of this kind  is incompatible with a holographic interpretation of a causal theory, so $\ell$ must be an infrared cutoff. Notice that in this case the non-relativistic limit in the field theory will be $c\to \infty$, $\ell\to\infty$ and $\kappa$ fixed, but an ultraviolet cutoff is still needed in order to have a relativistic completion. As we expected from field theory arguments, the $z<1$ theory can only be well defined in an intermediate range of scales.

The argument for Lifshitz geometries can easily be extended to any geometry with a local speed of light that decreases monotonically towards the boundary. Using the metric \eqref{eq:gaussmetric}, the variational probe for null geodesics has the Lagrangian
\beq
{\cal L}=-e^{2A+2B} \dot{t}^2 +e^{2A}\dot{x}^2+\dot{u}^2\,.
\eeq 
We can follow the same steps and introduce the conserved quantities $E$ and $P$
\beq
\dot{t}=E e^{-2 A-2B}, \quad \dot{x}=P e^{-2 A} \,, 
\eeq
which is followed by
\[
\dot{u}=e^{-A}\sqrt{E^2 e^{-2 B}-P^2}\,.
\]
Then the geodesic equations in terms of the radial coordinate are
\beq
{dt\over du}={E e^{-A-2B}\over \sqrt{E^2 e^{-2 B}-P^2}}\,, \quad  {dx\over du}={P e^{-A}\over \sqrt{E^2 e^{-2 B}-P^2}}\,.
\eeq
The local speed of light is simply $e^B$, so the argument of the square root will become negative at a finite value of $u$ for a null geodesic starting at the boundary. The arguments we have used for Lifshitz geometries hold for more general cases, geometries with such behavior all the way to the horizon are not sensible holographic duals.

\paragraph{Speed of light and the null energy condition.}

The null energy condition for Lifshitz geometries was studied in ref.~\cite{Koroteev:2007yp}. There it was shown that backgrounds with $z<1$ violate this condition, as opposed to backgrounds with $z\geq 1$. A natural question is then if the behavior of the local speed of light is related to the null energy condition so a consistent holographic description requires it. We will show that this is indeed the case.

We will use the metric \eqref{eq:gaussmetric} for the derivation. The NEC for a field theory implies that its energy-momentum tensor is semi-positive definite on the light cone
\beq\label{eq:NECgeneric}
T_{\mu\nu}\xi^\mu\xi^\nu \geq 0\,,
\eeq
for an arbitrary null vector $\xi^\mu$. For example for the perfect fluid this condition is transformed to $p+\rho\geq 0$, where $p$ and $\rho$ are pressure and energy density of the fluid, or if one introduces the equation of state $p=\omega\rho$, the condition implies $\omega\geq-1$, becoming exactly $\omega = -1$ for the cosmological constant. 

The energy-momentum tensor enters in the Einstein equations
\beq
R_{\mu\nu}-\frac{1}{2}g_{\mu\nu}R+\Lambda g_{\mu\nu} = T_{\mu\nu}\,,
\eeq
where $R_{\mu\nu}$ is the Ricci tensor and we have set the Newton's constant to be equal to one. Using the Einstein equations we can recast \eqref{eq:NECgeneric} in terms of the metric \eqref{eq:gaussmetric} and its derivatives only without specifying the matter content of the theory. As it was shown in \cite{Koroteev:2007yp}, for a diagonal energy-momentum tensor in presence of the spatial  $\text{SO}(D-2)$ isotropy \eqref{eq:NECgeneric} is equivalent to 
\[\label{eq:NECDiag}
R^t_t-R^x_x\leq 0\,,\quad R_t^t - R^u_u \leq 0\,,
\]
where the components of the Ricci tensor in the $D$ dimensional bulk read 
\<
R^t_t\eq -B''-D A' B'-B'^2-A''-(D-1)A'^2\,,\nln
R^x_x\eq   -A' B'-A''-(D-1)A'^2\,,\nln
R^u_u\eq -B''-\left(A'+B'\right)^2-(D-1) A''-(D-2) A'^2\,,
\>
where the primes stand for the derivatives with respect to $u$. 
The first inequality \eqref{eq:NECDiag} leads to
\[
B''+B'(B'+(D-1) A')\geq 0\,.
\]
Let us first see what happens when the null energy condition is satisfied. We define a function of the radial coordinate $C(u)$ such that
\[
B'=C e^{-(D-1)A-B}\,.
\]
Then the condition reads
\[
C'e^{-(D-1) A-B} \geq 0\,, 
\]
or merely $C'\geq 0$. The derivative of the local speed of light is
\[
(e^B)'=B'e^B=C e^{-(D-1) A}\,,
\]
whose sign is the same as the sign of $C$. Now we can study the different possibilities. First, if $C\geq 0$ for some value of $u=u_*$, then $C\geq 0$ for all $u>u_*$ and the speed of light is monotonically increasing with $u$. If instead $C<0$ for some value of $u=u_*$, then there will be a value $u_H>u_*$ such that $C(u_H)=0$ and we will go back to the first case. There could be a fine-tuned situation where $u_H\to \infty$, but it is unclear whether that is possible or not. 

If the null energy condition is violated, we can define the function $C(u)$, but now $C'<0$. Then, if $C<0$ for some value of $u=u_*$ the local speed of light will be monotonically decreasing for $u>u_*$. If, on the other hand, $C>0$ for some value of $u=u_*$, there will be a value $u_H>u_*$ such that $C(u_H)=0$. Again, there could be a fine-tuned situation where $u_H\to \infty$.

We have then proven that in generic situations the null energy condition is necessary in order to have a consistent holographic description. In particular,  for the Lifshitz metric \eqref{Metric:KLKLM} the NEC is satisfied if $z\geq 1$, only these values of $z$ are suitable for a holographic construction. Notice that there could be causality issues of the type we have described when the space is asymptotically AdS ($z=1$) if there is matter in the bulk that violates the NEC \cite{Gao:2000ga,Kleban:2001nh}.

For our considerations we do not need the second inequality from \eqref{eq:NECDiag} since the behavior of the local speed of light as a function of $u$ is governed by $tt$ and $xx$ components. In refs.~\cite{Freedman:1999gp,Girardello:1998pd,Myers:2010xs} the second condition from \eqref{eq:NECDiag} was used to model the RG running of central charges in a conformal field theory using a holography dual.

\section{Scalar two-point correlation functions.}\label{Sec:ScalarCorr}

We now study how the value of the dynamical exponent affects the correlation functions of operators in the field theory. We will consider scalar operators for simplicity, but we  expect that the main qualitative features will  be generic for operators of different spin. A scalar operator ${\cal O}_\Delta$ of scaling dimension $\Delta$ in the field theory is dual to a scalar field $\phi$ of mass $m^2L^2=\Delta(\Delta-d-z)$ in the  Lifshitz geometry \eqref{Metric:KLKLM}. The scalar correlator can be computed following the usual procedure of evaluating the action on a classical solution. The action is in general infinite, so it needs to be properly renormalized through the introduction of boundary counter-terms. We are not interested in the ultra-local behavior of the correlator, so we will ignore this issue and just keep the finite piece of the action. For a more complete discussion see ref.~\cite{Taylor:2008tg}. Euclidean correlation functions were originally computed for the $z=2$ case in ref.~\cite{Kachru:2008yh}. 

Given a plane-wave solution $\phi(t,\textbf{x},r)=e^{-i\omega t+i \textbf{k}\textbf{x}}\varphi_{\omega,\textbf{k}}(r)$ of the equations of motion, the holographic two-point function is
\beq\label{eq:boundaryaction}
G_2(\omega,\textbf{k})=-\lim_{\ell\to 0} \left.\sqrt{-g} g^{rr} \varphi_{\omega,\textbf{k}}'(r) \varphi_{\omega,\textbf{k}}(r)\right|_{r=\ell}\,.
\eeq
We have to specify boundary conditions for the solution in the $r\to \infty$ limit. The usual identification is that an ingoing wave corresponds to a retarded Green's function \cite{Son:2002sd}, however for $z<1$ it is not always possible to choose an ingoing condition. Instead, we will fix the solutions to be regular and when an ingoing solution is possible, to be the analytic continuation of a regular Euclidean solution. Therefore we will be computing Wightman correlators.
\beq
G_2(\omega,\textbf{k})=\int dt\, d^d \textbf{x}\, e^{i\omega t-i \textbf{k}\textbf{x}} \langle {\cal O}(t,\textbf{x}){\cal O}(0)\rangle\,.
\eeq

The equation of motion for a scalar field in \eqref{Metric:KLKLM} is
\beq\label{eq:scalareq}
\varphi''-{z+d-1\over r}\varphi'+{\omega^2\over \kappa^2} r^{2(z-1)}\varphi-k^2\varphi-{m^2\over r^2}\varphi = 0\,,
\]
where $k=|\textbf{\textbf{k}}|$. Notice that $\kappa$ has a mass dimension that depends on $z$, $[\kappa]=1-z$.

Let us now compute the two-point functions for three special values, $z=2$, $z=1$ and $z=1/2$ \footnote{The cases $z=0, z\to \infty$ can be found in ref.~\cite{Gordeli:2009vh}.}. For illustration purposes, it is enough if we consider massless fields $\Delta=d+z$ and fix the number of spatial dimensions to $d=2$. The solutions satisfying the appropriate boundary conditions are of the form $\varphi(r)=\Phi(r)/\Phi(\ell)$, where $\Phi(r)$ is the continuation of the regular Euclidean solution. We list our results in the following

\paragraph{z=2 (Lifshitz)} There is a subtlety here, the solution depends on whether we want to extend the correlator over the upper half of the frequency complex plane or over the lower part. The solution for ${\rm Im}\, \omega <0$ is
\[
\Phi_<(r)=e^{-{i\omega r^2\over 2 \kappa}} U\left(-{1\over 2}-{i \kappa k^2\over 4 \omega}\,,\,-1\,,\,  {i\omega r^2\over \kappa} \right)\,,
\]
where $U(a,b,c)$ is a confluent hypergeometric function. Meanwhile, the solution for ${\rm Im}\, \omega >0$ is
\[
\Phi_>(r)=e^{-{i\omega r^2\over 2 \kappa}}\left[L_{{1\over 2}+{i \kappa k^2\over 4 \omega}}^{\ \ 2}\left( {i\omega r^2\over \kappa}\right) +{i e^{k^2 \pi\over 4 \omega}\over \Gamma\left( {3\over 2}+{k^2 \pi\over 4 \omega}\right)} U\left(-{1\over 2}-{i \kappa k^2\over 4 \omega}\,,\,-1\,,\,  {i\omega r^2\over \kappa} \right)\right]\,,
\]
where $L^k_n(x)$ stands for a generalized Laguerre polynomial.

\paragraph{z=1 ($AdS$)}
\[
\Phi(r)=e^{-\sqrt{k^2-\omega^2} r} (1+\sqrt{k^2-\omega^2} r)\,.
\]

\paragraph{$z=\half$ (``Mirror'' Lifshitz)}

In this case the choice of the normalizable solution is unique and reads 
\[
\Phi(r)=r^{5/2}e^{-k r} \,U\left({7\over 4}-{\omega^2\over2 \kappa^2  k}\,,\,{7\over 2}\,,\,2 k r \right)\,.
\]
\vspace{3mm}

We now use \eqref{eq:boundaryaction} to compute the correlators, up to normalization factors and contact terms. In the $k\to 0$ limit the Schr\"oedinger equation  has the same form in all cases, with different coefficients in front of the $1/{r}^2$ potential \eqref{eq:Potentials}. The solutions are
\[
\Phi(r)=r^{1+{z\over 2}} K_{{1\over 2}+{1\over z}}\left({i|\omega| r^z\over z} \right)\,,
\]
where $K_\mu(x)$ is the modified Bessel function of the second kind. In the cases we study we find that
\begin{eqnarray}
\notag z=2, &  \Delta=4, & G_2(\omega,0)\simeq \omega^2\log(i\kappa\omega)\,,\\
\notag z=1, & \Delta=3, & G_2(\omega,0)\simeq |\omega|^3\,, \\
z={1\over 2}&  \Delta={5\over 2}, & G_2(\omega,0)\simeq |\omega|^5\,.
\end{eqnarray}
The behavior at non-zero momentum  is more interesting. Let us now calculate the two-point correlators

\paragraph{\textbf{z=2  ($\Delta=4$)}}
\[\label{eq:G2z2}
G_2(\omega,k)\simeq \left({4 \omega^2\over \kappa^2} +k^4\right) \left[\log\left(i\kappa \omega\right)+\psi\left({3\over 2}-{i \kappa k^2\over 4 \omega} \right)+i\Theta({\rm Im}\, \omega) \pi\,{\rm sech}\,\left({\kappa k^2\pi \over 4 \omega} \right)\right]\,,
\]
where $\psi(x)$ is the polygamma function and $\Theta(x)$ is the Heaviside step function. The correlator  has a branch cut along the positive imaginary axis. On top of it there are also poles at the positions
\[
\omega_n = { i\kappa k^2\over 4 n+6}\,, \quad n=0,1,2,\dots 
\]
There are two sources of poles in \eqref{eq:G2z2} -- the polygamma function and the ${\rm sech}$ function. For $n=2m-1\,, m=1,2,\dots$ both functions contribute, for the other $n$ only the $\psi$ has a pole giving a different residue at those poles.

Notice that there is an infinite set of poles at any vicinity of $\omega=0$. These poles all go to $\omega=0$ in the $k\to 0$ limit.

\paragraph{\textbf{$z=1$ ($\Delta=3$)}}
\begin{equation}
G_2(\omega,k)\simeq (k^2-\omega^2)^{3/2}\,.
\end{equation}
This is the usual relativistic result, with a branch cut for values of $\omega^2>k^2$.

\paragraph{\textbf{$z={1\over 2}$ ($\Delta={5\over 2}$)}}
\begin{equation}
G_2(\omega,k)\simeq  k^{5/2} {\Gamma\left( {7\over 4} -{\omega^2\over 2 k \kappa^2}\right)\over \Gamma\left( -{3\over 4} -{\omega^2\over 2 k \kappa^2}\right)}\,.
\end{equation}
In this case the correlator is analytic in the complex frequency plane. There is a set of discrete poles localized on the real axis
\begin{equation}\label{eq:poles}
\omega_n^2=\left( 2n+{7\over 2}\right) \kappa^2  k\,, \quad n=0,1,2,\dots
\end{equation}
and set of zeroes
\begin{equation}
\omega_n^2=\left( 2n-{3\over 2}\right) \kappa^2  k\,, \quad n=0,1,2,\dots
\end{equation}
When $k\to 0$, the poles and the zeroes go to $\omega=0$. The poles of the correlator give a gapless spectrum of propagating modes, with a dispersion relation $\omega \sim \sqrt{k}$. 

In order to show that these modes produce superluminal propagation we will follow the analysis in appendix C of ref.~\cite{Amado:2008ji}. A wavefront is produced by a source of the form $\Theta(t) e^{-i\nu t}\delta(x)$. In the linear approximation the response is approximately
\[
\langle {\cal O}(x,t)\rangle \simeq -\int {d \omega \over 2\pi} \sum_n R_n(\omega,k_n) {i\over \omega-\nu+i\epsilon} e^{-i\omega t + i k_n x} \,,
\]
where $k_n$ correspond to the modes \eqref{eq:poles} seen as poles in the complex {\em momentum} plane and $R_n$ are the residues. Specifically, 
\[
k_n =   {\omega^2\over \kappa^2\left( 2n+{7\over 2}\right) } \,, \quad n=0,1,2,\dots
\]
The phase velocity is defined as
\[
\left(v_{ph}\right)_n={\omega\over k_n} = {c\over \omega \ell} \left( 2n+{7\over 2}\right)\,.
\]
The properly defined wavefront velocity is obtained by taking the large frequency limit of the phase velocity defined as above. Naively it seems that $v_{ph}$ would vanish, but notice that for any given large value of the frequency $\omega\gg \ell/c$, there will always be an integer $n_\omega$ such that there is an infinite set of modes $n>n_\omega$ with $v_{ph} >c$. This implies that there is superluminal propagation, as we anticipated from the analysis of null geodesics.

\section{The NEC and Higher Derivative Gravity}\label{Sec:NECHighDer}

We have examined the relation between the null energy condition in the bulk and causality in the boundary theory for Lifshitz geometries and we have concluded that it should be satisfied in order to have a sensible holographic interpretation. A different kind of gravitational models that also produce Lifshitz geometries are higher derivative corrections of Einstein gravity, like the curvature squared corrections considered in ref.~\cite{Adams:2008zk,Cai:2009ac,AyonBeato:2010tm}. Higher derivative corrections of gravity and causality issues in holography have also been considered in the context of the AdS/CFT correspondence, mainly for Gauss-Bonnet gravity \cite{Brigante:2007nu,Brigante:2008gz,deBoer:2009pn,Buchel:2009tt,Buchel:2009sk,Hofman:2009ug,Camanho:2009vw,Ge:2008ni,Ge:2009eh,Ge:2009ac}, and also for quasi-topological gravity \cite{Myers:2010jv}. As we will see, our analysis is constrained to some particular class of models that in general do not include Gauss-Bonnet gravity, so we will not be able to impose further restrictions on this model.

In principle our analysis only shows that the NEC is violated for particular solutions, determined by the values of $\Lambda$ and the $\beta_i$ parameters. Notice that the cosmological constant $\Lambda$ does not enter in the NEC, so that for the values of $\beta_i$ where a solution with $z<1$ exists it is possible that the NEC is violated in more cases. For instance, Einstein gravity with a cosmological constant saturates the NEC, so if $R^2$ terms are introduced as small corrections, this implies that some mechanism should be at work to prevent general perturbations of the solutions from violating the NEC. A possibility could be to use boundary conditions such that problematic modes are projected out. This would require a thorough analysis of fluctuations, we will leave it as future work and take the simplest approach here. These arguments do not apply to Lifshitz solutions with $z>1$, as long as fluctuations are small they should not spoil the NEC. Notice that there is no contradiction here, even if the $R^2$ corrections are small, Lifshitz solutions have different asymptota to solutions to the Einstein equations, so they cannot be considered as small perturbations even if $z$ is close to one.

The authors of ref.~\cite{AyonBeato:2010tm} have made an extended analysis of various black hole solutions including Lifshitz black holes. The zero temperature Lifshitz metric is also a solution, and depending on the values of the parameters it is possible to find solutions with $z\geq 1$ or $z<1$. Whenever a solution with $z<1$ exists, this implies that the null energy condition can be violated for that choice of parameters.

A general form of the action is
\[\label{eq:GBlikeaction}
S = \int d^D x\sqrt{g} \left(R-2\Lambda+L^2\beta_1 R^2 +L^2\beta_2 R_{\alpha\beta}R^{\alpha\beta}+L^2\beta_3 R_{\alpha\beta\gamma\delta} R^{\alpha\beta\gamma\delta} \right)\,.
\]
The equations of motion read
\[\label{eq:EEGB}
R_{\mu\nu}-\frac{1}{2}g_{\mu\nu}R+\Lambda g_{\mu\nu}  = L^2 \Theta_{\mu\nu}\,,
\]
where $\Theta_{\mu\nu}$ stands for the variation of the higher derivative part of \eqref{eq:GBlikeaction} \cite{AyonBeato:2010tm}
$$
\Theta_{\mu\nu}={1\over 2} \left(\beta_1 R^2+\beta_2 R_{\alpha\beta} R^{\alpha\beta}+\beta_2 R_{\alpha\beta\gamma\delta} R^{\alpha\beta\gamma\delta} \right) g_{\mu\nu}
$$
$$
+4\beta_3 R_{\mu\alpha} R_{\nu}^\alpha-2\beta_1 R R_{\mu\nu}-2(\beta_2+2\beta_3) R_{\mu\alpha\nu\beta} R^{\alpha\beta}-2 \beta_3 R_{\mu\alpha\beta\gamma} R_\nu^{\alpha\beta\gamma}
$$
\[
+(2\beta_2+\beta_2+2 \beta_3) \nabla_\mu \nabla_\nu R-{1\over 2}(4\beta_1+\beta_2) g_{\mu\nu} \nabla^2 R-(\beta_2+4\beta_3) \nabla^2 R_{\mu\nu}\,.
\]
At the level of the equations of motion this tensor looks like some matter energy-momentum tensor that sources the Einstein equations. We can now look for Lifshitz solutions \eqref{Metric:KLKLM} of \eqref{eq:EEGB} which will partially fix the parameters $\Lambda$ and $\beta_i$ of the action \eqref{eq:GBlikeaction}.  For a general number of dimensions $D$, there are Lifshitz solutions provided that
\<
\label{eq:Lambda1}
\Lambda\eq -{1\over L^2}\Big[1+2(\beta_1-\beta_3)+2 z+\left(1-2 z+\half z^4\right)(4\beta_1+2\beta_2+4\beta_3)\nl
+(3 z^2-2 z^3)(\beta_2+4\beta_3) \Big]\,,
\>
\[
\label{eq:plane1}
2(2 z^2+(D-2)(2 z+D-1))\beta_1+2(z^2+D-2) \beta_2+4(z^2-(D-2)z+1)\beta_3=1\,.
\]
The latter condition determines a plane in the $(\beta_1,\beta_2,\beta_3)$ parameter space for each value of $z$. Notice that all planes intersect at a single point
\[\label{eq:specialpoint}
\tilde\beta_1=\tilde\beta_3=-\tilde\beta_2/4=1/(2(D-4)(D-3))\,.
\]
These values correspond to a particular case of Gauss-Bonnet gravity. Here and in the following  we will assume that $D>4$. If we solve for $\beta_1$ or $\beta_2$ in \eqref{eq:plane1} and plug the result in \eqref{eq:Lambda1}, we find
\[\label{eq:LambdaR2}
\Lambda=-{1\over 4 L^2}\left(2z+(D-2)(2 z+D-1)-4(D-3)(D-4)z(z+D-2)\beta_3\right) \,.
\]

There is a second branch of solutions with $z=1$, where the cosmological constant is fixed in terms of the $\beta_i$'s \eqref{eq:Lambda1} but there is no constraint \eqref{eq:plane1}. Notice that the two branches will have the same value of parameters only when a $z=1$ solution is allowed in the first branch, so the cosmological constant is the same. This happens at the points in the plane
\[
\label{eq:plane2}
2D(D-1)\beta_1+2(D-1) \beta_2-4(D-4)\beta_3=1\,.
\]

The condition for the existence of Lifshitz solutions \eqref{eq:plane1} can be written as a homogeneous equation for planes centered around the special point \eqref{eq:specialpoint}
\[
a_1(z)(\beta_1-\tilde\beta_1)+a_2(z)(\beta_2-\tilde\beta_2)+a_3(z)(\beta_3-\tilde\beta_3)=0\,,
\]
where
\begin{eqnarray}\label{eq:coeffs}
\notag a_1(z) & = & 2(2 z^2+(D-2)(2 z+D-1)),\\
\notag a_2(z) & = & 2(z^2+D-2), \\ 
a_3(z)& = & 4(z^2-(D-2)z+1).
\end{eqnarray}

The set of planes defined by the values $z\in(-\infty,\infty)$ span a volume in the $(\beta_1,\beta_2,\beta_3)$ space. Models with $R^2$ corrections to Einstein-Hilbert gravity where violations of the null energy condition is possible lie in the region $-\infty<z<1$. Notice that the value of the cosmological constant is different for models with different values of $z$ and the same values of the $\beta_i$ coefficients. 

\begin{figure}[htb!]
\begin{center}
\begin{tabular}{cc}
\includegraphics[width=7cm]{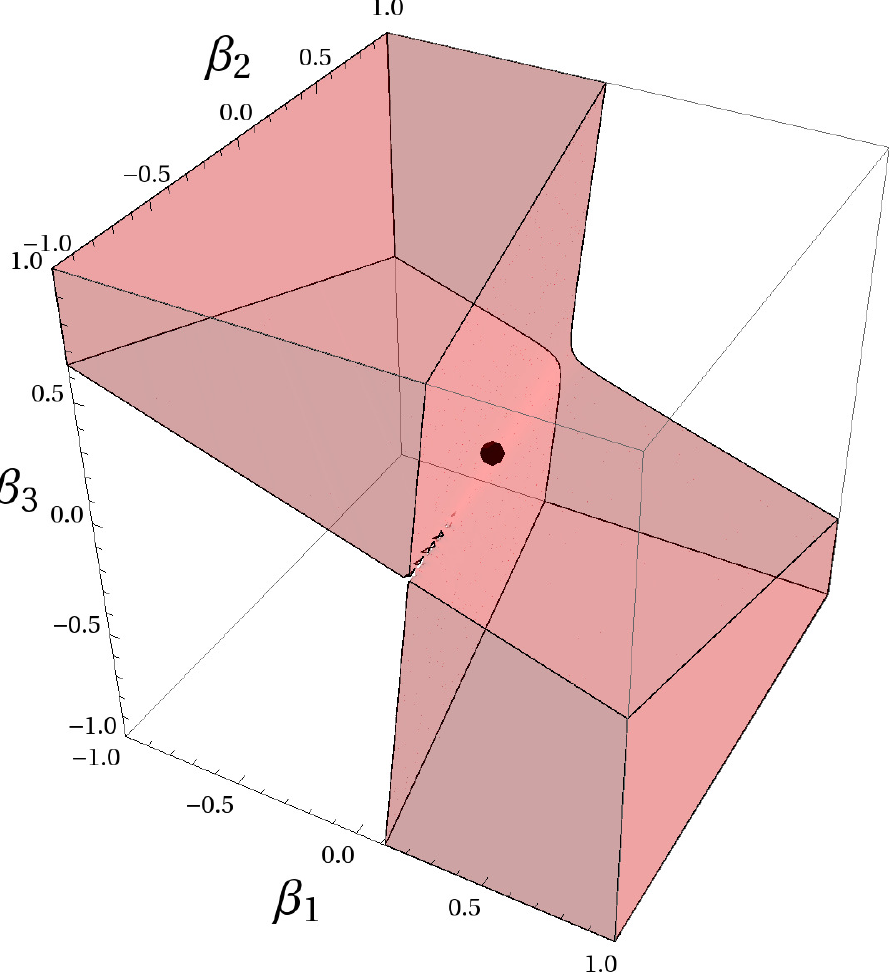} &
\includegraphics[width=7cm]{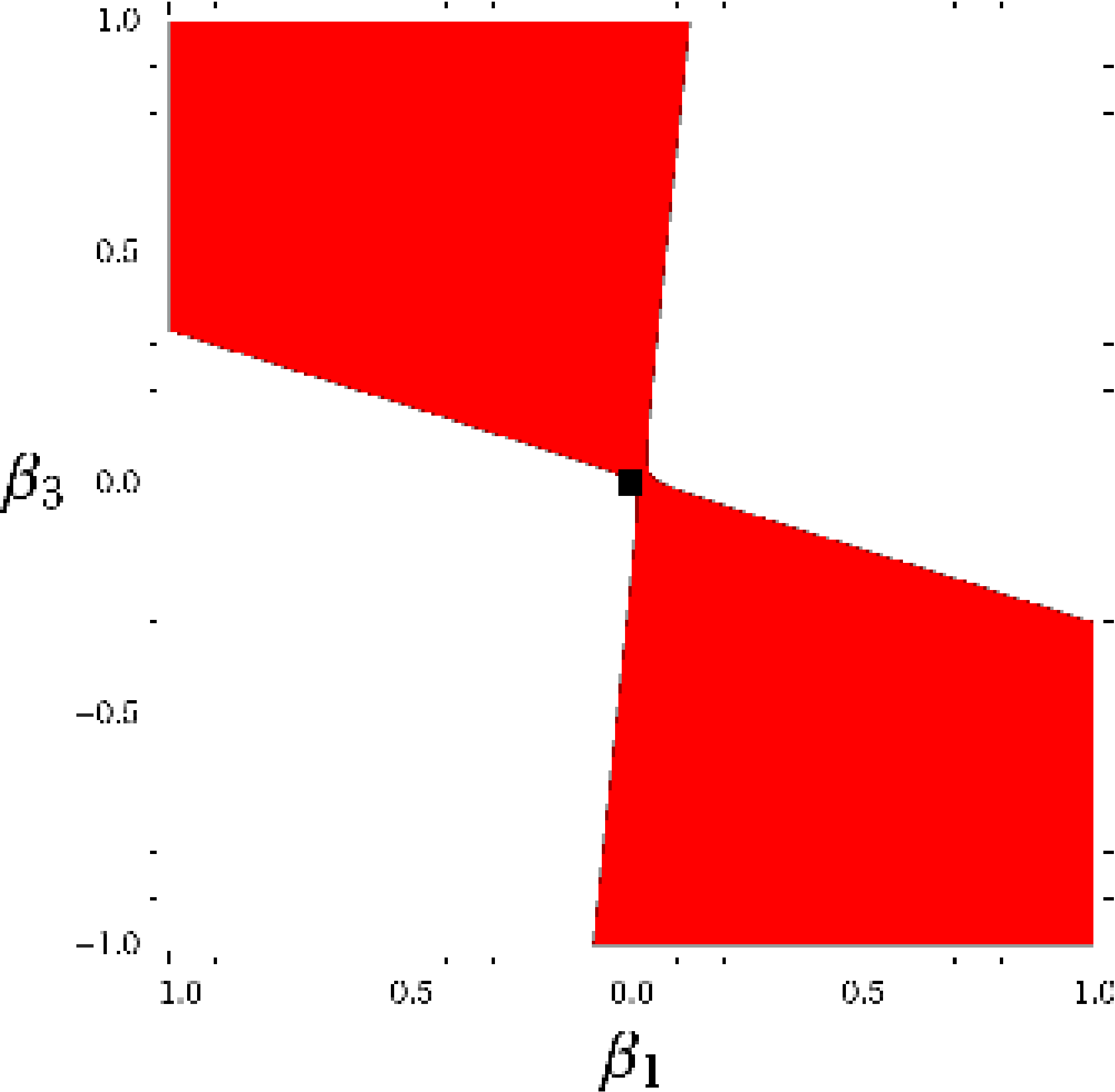}
\end{tabular}
\caption{\label{fig:domains}  Solutions with Lifshitz scaling are allowed in the colored regions of $(\beta_1,\beta_2,\beta_3)$ space (left). Solutions with $z<1$  exist in the full region for determined values of the cosmological constant, so violations of the NEC are possible in the full region. The thick dot in the origin corresponds to Einstein gravity. The right plot is a slice on the $\beta_2=0$ plane, where the dot corresponds to Einstein gravity, we observe that it lies at the boundary of the Lifshitz region, where two different $z=1$ solutions exist.}

\end{center}
\end{figure}

Given a value of $z$ the theories that admit a Lifshitz solution lie on a plane that contains the origin of the $x_i=\beta_i-\tilde\beta_i$ space. The plane is determined by the equation ${\bf a}(z)\cdot {\bf x} =0$, where we use bold face for three-dimensional vectors. If we shift the value of the dynamical exponent $z\to z+\delta z$, Lifshitz solutions will be allowed in the plane  ${\bf a}(z+\delta z)\cdot {\bf x}=0$. For an infinitesimal variation ${\bf a}(z+\delta z)\simeq {\bf a}(z)+{\bf a}'(z)\delta z$ the intersection between the two planes is given by the line
\[
\label{eq:line}
{\bf a}(z)\cdot {\bf x}=0, \ \ \ {\bf a}'(z)\cdot {\bf x}=0.
\]
A general solution is given parametrically as
\begin{eqnarray}
\notag X_1 & = & \tau (a_2(z) a_3'(z)-a_2'(z) a_3(z)), \\
\notag X_2 & = & \tau (a_3(z) a_1'(z)-a_3'(z) a_1(z)), \\
X_3 & = & \tau (a_1(z) a_2'(z)-a_1'(z) a_2(z)).
\end{eqnarray}
As we vary $z$, the lines span a surface, the envelope of the planes. The explicit form is
\begin{eqnarray}\label{eq:surface}
\notag X_1 & = & 8\tau \left((D-2) z^2+2(D-3) z-(D-2)^2\right), \\
\notag X_2 & = & -8\tau\left(4(D-2) z^2+2 D(D-3) z-(D-2)(D(D-3)+4) \right), \\
X_3 & = & 8\tau\left((D-2) z^2+(D-2)(D-3) z-(D-2)^2\right).
\end{eqnarray}
Each plane is tangent to this surface along the line \eqref{eq:line}. Furthermore, the equation ${\bf a}(z_1)\cdot {\bf X}(z_2,\tau)=0$ is satisfied only if $z_1=z_2$, so each plane intersects the surface ${\bf X}$ only once. We can solve $z$ as a function of $x_i$ as well, there are two possible solutions
$$
z_\pm={1\over 2 x_1+x_2+2 x_3}\left[(D-2) (x_3-x_1)\pm\right.
$$
\[\label{eq:eqz}
\left.
\pm \sqrt{(D-2)^2(x_1-x_3)^2-(2 x_1+x_2+2 x_3) ((D-2)(D-1)x_1+(D-2) x_2+2 x_3)} \right]\,.
\]
The region with allowed Lifshitz solutions corresponds to values of $x_i$ such that the argument of the square root is positive. The boundary of the allowed region is represented by the surface
\[\label{eq:implicit}
(D-2)^2(x_1-x_3)^2-(2 x_1+x_2+2 x_3) ((D-2)(D-1)x_1+(D-2) x_2+2 x_3)=0.
\]
This is just the implicit form of the envelope surface \eqref{eq:surface}, as one can easily check by introducing the explicit expressions in the equation \eqref{eq:implicit} with $x_i=X_i$. In order to avoid Lifshitz solutions with $z<1$ we should impose $z_+>1$ and $z_->1$. This can be simplified if we impose first the condition $z_+ z_->1$, then discard negative values of $z$ and finally discard the cases where 
\[\label{eq:ExtraCondition}
z_\pm >\frac{1}{z_\mp} >1\,.
\]
We find that
\[
z_+ z_- = {(D-2)(D-1)x_1+(D-2) x_2+2 x_3\over 2 x_1+x_2+2 x_3} >1\,,
\]
then
\[
\begin{array}{ccc} 
((D-2)(D-1)-2) x_1+(D-3) x_2 >0 & {\rm if} & 2 x_1+x_2+2 x_3>0\,, \\
((D-2)(D-1)-2) x_1+(D-3) x_2 <0 & {\rm if} & 2 x_1+x_2+2 x_3<0\,, 
\end{array}
\]
and since $z_+$ and $z_-$ have to be positive, we also have
\[
\begin{array}{ccc} 
(D-2)(D-1) x_1+(D-2) x_2 +2 x_3>0 & {\rm if} & 2 x_1+x_2+2 x_3>0\,, \\
(D-2)(D-1) x_1+(D-2) x_2 +2 x_3<0 & {\rm if} & 2 x_1+x_2+2 x_3<0\,. 
\end{array}
\]
Notice that in this region the argument inside the square root in \eqref{eq:eqz} is always smaller in absolute value than the first term squared, since  $(2 x_1+x_2+2 x_3) ((D-2)(D-1)x_1+(D-2) x_2+2 x_3)>0$. The regions where $z_\pm$ are positive are
\[
\begin{array}{ccc}\label{eq:positive}
x_3 > x_1 & {\rm if} & 2 x_1+x_2+2 x_3 >0\,, \\ 
x_3 < x_1 & {\rm if} & 2 x_1+x_2+2 x_3 <0\,. 
\end{array}
\]
As far as \eqref{eq:ExtraCondition} is concerned, we can write $z_\pm=(A\pm B)/C$, where $A=(D-2)(x_3-x_1)$, $B$ is the square root term in \eqref{eq:eqz} and $C=2 x_1+x_2+2 x_3$. From \eqref{eq:positive} we have the conditions that $B\geq 0$ and $A$ and $C$ have the same sign. Let us first consider the case $C\geq 0$, $A\geq 0$. Clearly $z_+\geq z_-$, now imagine that we are in the situation we want to discard, when $z_+\geq 1\geq z_-$. We can do the following manipulations
\[
\left.
\begin{array}{ccc}
z_+={A+B\over C} \geq 1, & \Rightarrow & A+B\geq C \\
z_-={A-B\over C} \leq 1, & \Rightarrow & A-B\leq C
\end{array}
\right\}\Rightarrow B\geq 0\,. 
\]
We know that the last expression is true, so the only allowed possibility is $z_+=z_-=1$. The case $C\leq 0$, $A\leq 0$ works in a similar way, if we consider the case $z_-\geq 1\geq z_+$,
\[
\left.
\begin{array}{ccc}
z_+={A+B\over C} \leq 1, & \Rightarrow & A+B\geq C \\
z_-={A-B\over C} \geq 1, & \Rightarrow & A-B\leq C
\end{array}
\right\}\Rightarrow B\geq 0 \,,
\]
which again is always true. Therefore the NEC can be violated in the full region of parameter space where Lifshitz solutions with $z\neq 1$ exist. Our results for the case $D=5$ are summarized in \figref{fig:domains}.

\section{Conclusions and Open Questions}\label{Sec:Conclusions}

In this paper we have investigated the role of the null energy condition in a gravitational model with respect to causality in holography. We have shown that the NEC  has a simple physical manifestation in terms of the local speed of light in the bulk. If the local speed of light on a radial slice increases when an observer moves towards the boundary, then and only then the NEC is satisfied. If we consider the Lifshitz metric with the critical exponent $z$ \eqref{Metric:KLKLM}, the NEC merely requires $z\geq 1$. The behavior of the speed of light when the NEC is violated implies that shock wave sources can bounce back to the boundary and reach it at points that are outside the light cone of signals propagating along the boundary. The front of radiation produced by the shock wave will also move at superluminal speeds on the average. Based on the results of ref.~\cite{Amado:2008hw}, we have argued that this would lead to superluminal modes, so we have continued with the investigation of the spectra of scalar field perturbations in Lifshitz backgrounds for different values of the critical exponent. We have shown explicitly that for a Lifshitz background with $z=1/2$ there is indeed a set of discrete modes whose dispersion relation is incompatible with causality, while for examples with $z>1$ such modes are absent. We have related the presence of such modes to the properties of the equation of motion for backgrounds with $z<1$, by showing that the equation written in the Schr\"{o}dinger form has a confining potential at non-zero momentum.

Summarizing, we have presented strong arguments and some evidence that geometries produced by matter that violates the NEC will produce superluminal propagation in the dual theory. This is directly related to the observation in higher derivative theories that fluctuations in AdS with negative energy fluxes will spoil causality in the CFT \cite{Brigante:2007nu,Brigante:2008gz,deBoer:2009pn,Buchel:2009tt,Buchel:2009sk,Hofman:2009ug,Camanho:2009vw}.\footnote{A recent work also connects negative energy fluxes to the presence of ghosts in the CFT \cite{{Kulaxizi:2010jt}}} It would be interesting to confirm this statement in more general cases. There is also still an open question of whether violations of the NEC in a localized region of the bulk, like those produced by quantum effects, are allowed in holographic systems. It is also interesting to observe that the NEC seems to be an important condition for several aspects of holography, in principle not directly related to each other in the field theory. In addition to causality constraints, it allows the formulation of holographic c-theorems \cite{Freedman:1999gp,Girardello:1998pd,Myers:2010xs} and it was originally proposed in the context of the entropy bound \cite{Bousso:1999xy,Bousso:2002ju}.
A possible application of these results is to constrain extensions of holographic models with new forms of matter or curvature corrections to Einstein gravity. We have done this for holographic models that involve curvature squared corrections to Einstein gravity. Our criterion was to discard the regions of the parameter space that allow Lifshitz solutions with $z<1$ and hence violations of the NEC by small perturbations around the solutions to Einstein equations. 
More generally, one would constrain possible holographic constructions by separating Einstein equations in two parts -- one being the Einstein tensor and the other an effective energy-momentum tensor that includes the higher curvature corrections and any additional matter and imposing the null energy condition on the last.

We have worked mainly in the bulk theory, let us now draw our attention to the macroscopical properties of the boundary theory. We can consider introducing a finite temperature in the Lifshitz theory, or in the dual description a black hole  \cite{Danielsson:2009gi, Brynjolfsson:2010mk,Bertoldi:2009vn}. Given the scale invariance of the theory, the equation of state for a perfect fluid should have the form
\[\label{eq:IsotropyBoundary}
z\langle T_{tt} \rangle -d \langle T_{xx} \rangle =0\,.
\]
Assuming the metric of the boundary theory is flat let us consider the so-called dominant energy condition (DEC) for the boundary theory\footnote{Boundary NEC is automatically satisfied for positive energy density.}. The DEC is responsible for the causal flow of the energy-momentum. In covariant formulation is states 
\[
-T^0_\nu\, \xi^\nu >0\,,
\]
for any light-like or timelike vector $\xi^\mu$ such that $\xi^0>0$. For the theory with spatial isotropy the DEC reads
\[\label{eq:DECboundary}
\langle T_{tt}\rangle\geq 0, \ \ \ \langle T_{tt}\rangle -|\langle T_{xx}\rangle| =\langle T_{tt}\rangle\left(1-{z\over d} \right)\geq 0\,,
\]
where we used \eqref{eq:IsotropyBoundary}. The boundary DEC is satisfied provided $z \leq d$. This condition, combined together with the constraint from causality we have discussed in \secref{Sec:LocalSpeed} gives the domain of critical exponents
\[\label{eq:Allowedz}
1\leq z \leq d\,.
\]
Notice that the upper bound has been derived from the boundary DEC, while the lower
bound has been derived from the bulk NEC. It is not clear how seriously the upper bound on the critical exponent should be considered. For instance, the supergravity construction of \cite{Donos:2010ax} allows for a Lifshitz background with $z\approx 39$; even real condensed matter systems \cite{Si:2001kx} with $z>d$ are known. The most likely explanation, at least for the condensed matter systems, is that the DEC cannot be applied to the energy-momentum tensor we are considering. 

\section*{Acknowledgements}

We would like to thank J. Gauntlett, A. Vainstein, S. Hartnoll, A. Karch, V. Lysov, and R. Myers for fruitful discussions, and A\&M University in College Station, TX for hospitality. P.K. is grateful to Institute d'\'Etudes Scientifiques de Cargese, France for kind hospitality, where some part of his work was done. This work was supported in part by the U.S. Department of Energy under Grants No. DE-FG02-94ER40823 (P.K.) and No. DE-FG02-96ER40956. (C.H.)


\bibliography{holliv}

\begin{thebibliography}{10}
\ifx\href\asklfhas\newcommand{\href}[2]{#2}\fi
\ifx\arxivref\asklfhas\newcommand{\arxivref}[1]{\href{http://arxiv.org/abs/#1}%
{#1}}\fi
\ifx\doiref\asklfhas\newcommand{\doiref}[2]{\href{http://dx.doi.org/#1}{#2}}\fi
\raggedright
\small
\parskip 0pt

\bibitem{Maldacena:1997re}
J.~M.~Maldacena,
\textit{``{The large N limit of superconformal field theories and
  supergravity}''},
\textsf{Adv.~Theor.~Math.~Phys.~2,~231~(1998)},
\texttt{\arxivref{hep-th/9711200}}.
%
\bibitem{Gubser:1998bc}
S.~S.~Gubser, I.~R.~Klebanov and A.~M.~Polyakov,
\textit{``{Gauge theory correlators from non-critical string theory}''},
\textsf{\doiref{10.1016/S0370-2693(98)00377-3}{Phys.~Lett.~B428,~105~(1998)}},
\texttt{\arxivref{hep-th/9802109}}.
%
\bibitem{Hartnoll:2009sz}
S.~A.~Hartnoll,
\textit{``{Lectures on holographic methods for condensed matter physics}''},
\textsf{\doiref{10.1088/0264-9381/26/22/224002}{Class.~Quant.~Grav.~26,~224002%
~(2009)}},
\texttt{\arxivref{0903.3246}}.
%
\bibitem{Herzog:2009xv}
C.~P.~Herzog,
\textit{``{Lectures on Holographic Superfluidity and Superconductivity}''},
\textsf{\doiref{10.1088/1751-8113/42/34/343001}{J.~Phys.~A42,~343001~(2009)}},
\texttt{\arxivref{0904.1975}}.
%
\bibitem{Lifshitz:1941aa}
E.~M.~Lifshitz,
\textit{``{On the Theory of Second-Order Phase Transitions I \& II}''},
\textsf{Zh.~Eksp.~Teor.~Fiz.~11,~255~(1941)}.
%
\bibitem{Son:2008ye}
D.~T.~Son,
\textit{``{Toward an AdS/cold atoms correspondence: a geometric realization of
  the Schroedinger symmetry}''},
\textsf{\doiref{10.1103/PhysRevD.78.046003}{Phys.~Rev.~D78,~046003~(2008)}},
\texttt{\arxivref{0804.3972}}.
%
\bibitem{Kachru:2008yh}
S.~Kachru, X.~Liu and M.~Mulligan,
\textit{``{Gravity Duals of Lifshitz-like Fixed Points}''},
\textsf{\doiref{10.1103/PhysRevD.78.106005}{Phys.~Rev.~D78,~106005~(2008)}},
\texttt{\arxivref{0808.1725}}.
%
\bibitem{Adams:2008wt}
A.~Adams, K.~Balasubramanian and J.~McGreevy,
\textit{``{Hot Spacetimes for Cold Atoms}''},
\textsf{\doiref{10.1088/1126-6708/2008/11/059}{JHEP~0811,~059~(2008)}},
\texttt{\arxivref{0807.1111}}.
%
\bibitem{1991RSPSA.434....9K}
A.~N.~{Kolmogorov},
\textit{``{The local structure of turbulence in incompressible viscous fluid
  for very large Reynolds numbers}''},
\textsf{Royal~Society~of~London~Proceedings~Series~A~434,~9~(1991)}.
%
\bibitem{Koroteev:2007yp}
P.~Koroteev and M.~Libanov,
\textit{``{On Existence of Self-Tuning Solutions in Static Braneworlds without
  Singularities}''},
\textsf{\doiref{10.1088/1126-6708/2008/02/104}{JHEP~0802,~104~(2008)}},
\texttt{\arxivref{0712.1136}}.
%
\bibitem{Gordeli:2009vh}
I.~Gordeli and P.~Koroteev,
\textit{``{Comments on Holography with Broken Lorentz Invariance}''},
\textsf{\doiref{10.1103/PhysRevD.80.126001}{Phys.~Rev.~D80,~126001~(2009)}},
\texttt{\arxivref{0904.0509}}.
%
\bibitem{Dubovsky:2001fj}
S.~L.~Dubovsky,
\textit{``Tunneling into extra dimension and high-energy violation of Lorentz
  invariance''},
\textsf{JHEP~0201,~012~(2002)},
\texttt{\arxivref{hep-th/0103205}}.
%
\bibitem{Koroteev:2009xd}
P.~Koroteev and M.~Libanov,
\textit{``{Spectra of Field Fluctuations in Braneworld Models with Broken Bulk
  Lorentz Invariance}''},
\textsf{\doiref{10.1103/PhysRevD.79.045023}{Phys.~Rev.~D79,~045023~(2009)}},
\texttt{\arxivref{0901.4347}}.
%
\bibitem{Balasubramanian:1998de}
V.~Balasubramanian, P.~Kraus, A.~E.~Lawrence and S.~P.~Trivedi,
\textit{``{Holographic probes of anti-de Sitter space-times}''},
\textsf{\doiref{10.1103/PhysRevD.59.104021}{Phys.~Rev.~D59,~104021~(1999)}},
\texttt{\arxivref{hep-th/9808017}}.
%
\bibitem{Klebanov:1999tb}
I.~R.~Klebanov and E.~Witten,
\textit{``{AdS/CFT correspondence and symmetry breaking}''},
\textsf{\doiref{10.1016/S0550-3213(99)00387-9}{Nucl.~Phys.~B556,~89~(1999)}},
\texttt{\arxivref{hep-th/9905104}}.
%
\bibitem{Landau:1991}
L.~D.~Landau and E.~M.~Lifshitz,
\textit{``Quantum Mechanics: Non-relativistic Theory, vol. 3''},
Butterworth-Heinemann (1991),
677p.
%
\bibitem{Horowitz:1995gi}
G.~T.~Horowitz and D.~Marolf,
\textit{``{Quantum probes of space-time singularities}''},
\textsf{\doiref{10.1103/PhysRevD.52.5670}{Phys.~Rev.~D52,~5670~(1995)}},
\texttt{\arxivref{gr-qc/9504028}}.
%
\bibitem{Gubser:2009qt}
S.~S.~Gubser and F.~D.~Rocha,
\textit{``{Peculiar properties of a charged dilatonic black hole in
  $AdS_5$}''},
\textsf{\doiref{10.1103/PhysRevD.81.046001}{Phys.~Rev.~D81,~046001~(2010)}},
\texttt{\arxivref{0911.2898}}.
%
\bibitem{Goldstein:2009cv}
K.~Goldstein, S.~Kachru, S.~Prakash and S.~P.~Trivedi,
\textit{``{Holography of Charged Dilaton Black Holes}''},
\texttt{\arxivref{0911.3586}}.
%
\bibitem{Hartnoll:2009ns}
S.~A.~Hartnoll, J.~Polchinski, E.~Silverstein and D.~Tong,
\textit{``{Towards strange metallic holography}''},
\texttt{\arxivref{0912.1061}}.
%
\bibitem{Azeyanagi:2009pr}
T.~Azeyanagi, W.~Li and T.~Takayanagi,
\textit{``{On String Theory Duals of Lifshitz-like Fixed Points}''},
\texttt{\arxivref{0905.0688}}.
%
\bibitem{Horava:2009vy}
P.~Horava and C.~M.~Melby-Thompson,
\textit{``{Anisotropic Conformal Infinity}''},
\texttt{\arxivref{0909.3841}}.
%
\bibitem{Guica:2010sw}
M.~Guica, K.~Skenderis, M.~Taylor and B.~van~Rees,
\textit{``{Holography for Schrodinger backgrounds}''},
\texttt{\arxivref{1008.1991}}.
%
\bibitem{Hofman:2008ar}
D.~M.~Hofman and J.~Maldacena,
\textit{``{Conformal collider physics: Energy and charge correlations}''},
\textsf{\doiref{10.1088/1126-6708/2008/05/012}{JHEP~0805,~012~(2008)}},
\texttt{\arxivref{0803.1467}}.
%
\bibitem{Chesler:2008uy}
P.~M.~Chesler, K.~Jensen, A.~Karch and L.~G.~Yaffe,
\textit{``{Light quark energy loss in strongly-coupled N = 4 supersymmetric
  Yang-Mills plasma}''},
\textsf{\doiref{10.1103/PhysRevD.79.125015}{Phys.~Rev.~D79,~125015~(2009)}},
\texttt{\arxivref{0810.1985}}.
%
\bibitem{Chesler:2008wd}
P.~M.~Chesler, K.~Jensen and A.~Karch,
\textit{``{Jets in strongly-coupled N = 4 super Yang-Mills theory}''},
\textsf{\doiref{10.1103/PhysRevD.79.025021}{Phys.~Rev.~D79,~025021~(2009)}},
\texttt{\arxivref{0804.3110}}.
%
\bibitem{Hatta:2008tx}
Y.~Hatta, E.~Iancu and A.~H.~Mueller,
\textit{``{Jet evolution in the N=4 SYM plasma at strong coupling}''},
\textsf{\doiref{10.1088/1126-6708/2008/05/037}{JHEP~0805,~037~(2008)}},
\texttt{\arxivref{0803.2481}}.
%
\bibitem{Amado:2008hw}
I.~Amado and C.~Hoyos-Badajoz,
\textit{``{AdS black holes as reflecting cavities}''},
\textsf{\doiref{10.1088/1126-6708/2008/09/118}{JHEP~0809,~118~(2008)}},
\texttt{\arxivref{0807.2337}}.
%
\bibitem{Gao:2000ga}
S.~Gao and R.~M.~Wald,
\textit{``{Theorems on gravitational timed related issues}''},
\textsf{\doiref{10.1088/0264-9381/17/24/305}{Class.~Quant.~Grav.~17,~4999~(200%
0)}},
\texttt{\arxivref{gr-qc/0007021}}.
%
\bibitem{Kleban:2001nh}
M.~Kleban, J.~McGreevy and S.~D.~Thomas,
\textit{``{Implications of bulk causality for holography in AdS}''},
\textsf{JHEP~0403,~006~(2004)},
\texttt{\arxivref{hep-th/0112229}}.
%
\bibitem{Freedman:1999gp}
D.~Z.~Freedman, S.~S.~Gubser, K.~Pilch and N.~P.~Warner,
\textit{``{Renormalization group flows from holography supersymmetry and a
  c-theorem}''},
\textsf{Adv.~Theor.~Math.~Phys.~3,~363~(1999)},
\texttt{\arxivref{hep-th/9904017}}.
%
\bibitem{Girardello:1998pd}
L.~Girardello, M.~Petrini, M.~Porrati and A.~Zaffaroni,
\textit{``{Novel local CFT and exact results on perturbations of N = 4 super
  Yang-Mills from AdS dynamics}''},
\textsf{JHEP~9812,~022~(1998)},
\texttt{\arxivref{hep-th/9810126}}.
%
\bibitem{Myers:2010xs}
R.~C.~Myers and A.~Sinha,
\textit{``{Seeing a c-theorem with holography}''},
\texttt{\arxivref{1006.1263}}.
%
\bibitem{Taylor:2008tg}
M.~Taylor,
\textit{``{Non-relativistic holography}''},
\texttt{\arxivref{0812.0530}}.
%
\bibitem{Son:2002sd}
D.~T.~Son and A.~O.~Starinets,
\textit{``{Minkowski-space correlators in AdS/CFT correspondence: Recipe and
  applications}''},
\textsf{JHEP~0209,~042~(2002)},
\texttt{\arxivref{hep-th/0205051}}.
%
\bibitem{Amado:2008ji}
I.~Amado, C.~Hoyos-Badajoz, K.~Landsteiner and S.~Montero,
\textit{``{Hydrodynamics and beyond in the strongly coupled N=4 plasma}''},
\textsf{\doiref{10.1088/1126-6708/2008/07/133}{JHEP~0807,~133~(2008)}},
\texttt{\arxivref{0805.2570}}.
%
\bibitem{Adams:2008zk}
A.~Adams, A.~Maloney, A.~Sinha and S.~E.~Vazquez,
\textit{``{1/N Effects in Non-Relativistic Gauge-Gravity Duality}''},
\textsf{\doiref{10.1088/1126-6708/2009/03/097}{JHEP~0903,~097~(2009)}},
\texttt{\arxivref{0812.0166}}.
%
\bibitem{Cai:2009ac}
R.-G.~Cai, Y.~Liu and Y.-W.~Sun,
\textit{``{A Lifshitz Black Hole in Four Dimensional R2 Gravity}''},
\textsf{\doiref{10.1088/1126-6708/2009/10/080}{JHEP~0910,~080~(2009)}},
\texttt{\arxivref{0909.2807}}.
%
\bibitem{AyonBeato:2010tm}
E.~Ayon-Beato, A.~Garbarz, G.~Giribet and M.~Hassaine,
\textit{``{Analytic Lifshitz black holes in higher dimensions}''},
\textsf{\doiref{10.1007/JHEP04(2010)030}{JHEP~1004,~030~(2010)}},
\texttt{\arxivref{1001.2361}}.
%
\bibitem{Brigante:2007nu}
M.~Brigante, H.~Liu, R.~C.~Myers, S.~Shenker and S.~Yaida,
\textit{``{Viscosity Bound Violation in Higher Derivative Gravity}''},
\textsf{\doiref{10.1103/PhysRevD.77.126006}{Phys.~Rev.~D77,~126006~(2008)}},
\texttt{\arxivref{0712.0805}}.
%
\bibitem{Brigante:2008gz}
M.~Brigante, H.~Liu, R.~C.~Myers, S.~Shenker and S.~Yaida,
\textit{``{The Viscosity Bound and Causality Violation}''},
\textsf{\doiref{10.1103/PhysRevLett.100.191601}{Phys.~Rev.~Lett.~100,~191601~(%
2008)}},
\texttt{\arxivref{0802.3318}}.
%
\bibitem{deBoer:2009pn}
J.~de~Boer, M.~Kulaxizi and A.~Parnachev,
\textit{``{$AdS_7/CFT_6$, Gauss-Bonnet Gravity, and Viscosity Bound}''},
\textsf{\doiref{10.1007/JHEP03(2010)087}{JHEP~1003,~087~(2010)}},
\texttt{\arxivref{0910.5347}}.
%
\bibitem{Buchel:2009tt}
A.~Buchel and R.~C.~Myers,
\textit{``{Causality of Holographic Hydrodynamics}''},
\textsf{\doiref{10.1088/1126-6708/2009/08/016}{JHEP~0908,~016~(2009)}},
\texttt{\arxivref{0906.2922}}.
%
\bibitem{Buchel:2009sk}
A.~Buchel et~al.,
\textit{``{Holographic GB gravity in arbitrary dimensions}''},
\textsf{\doiref{10.1007/JHEP03(2010)111}{JHEP~1003,~111~(2010)}},
\texttt{\arxivref{0911.4257}}.
%
\bibitem{Hofman:2009ug}
D.~M.~Hofman,
\textit{``{Higher Derivative Gravity, Causality and Positivity of Energy in a
  UV complete QFT}''},
\textsf{\doiref{10.1016/j.nuclphysb.2009.08.001}{Nucl.~Phys.~B823,~174~(2009)}%
},
\texttt{\arxivref{0907.1625}}.
%
\bibitem{Camanho:2009vw}
X.~O.~Camanho and J.~D.~Edelstein,
\textit{``{Causality constraints in AdS/CFT from conformal collider physics and
  Gauss-Bonnet gravity}''},
\textsf{\doiref{10.1007/JHEP04(2010)007}{JHEP~1004,~007~(2010)}},
\texttt{\arxivref{0911.3160}}.
%
\bibitem{Ge:2008ni}
X.-H.~Ge, Y.~Matsuo, F.-W.~Shu, S.-J.~Sin and T.~Tsukioka,
\textit{``{Viscosity Bound, Causality Violation and Instability with Stringy
  Correction and Charge}''},
\textsf{\doiref{10.1088/1126-6708/2008/10/009}{JHEP~0810,~009~(2008)}},
\texttt{\arxivref{0808.2354}}.
%
\bibitem{Ge:2009eh}
X.-H.~Ge and S.-J.~Sin,
\textit{``{Shear viscosity, instability and the upper bound of the Gauss-Bonnet
  coupling constant}''},
\textsf{\doiref{10.1088/1126-6708/2009/05/051}{JHEP~0905,~051~(2009)}},
\texttt{\arxivref{0903.2527}}.
%
\bibitem{Ge:2009ac}
X.-H.~Ge, S.-J.~Sin, S.-F.~Wu and G.-H.~Yang,
\textit{``{Shear viscosity and instability from third order Lovelock
  gravity}''},
\textsf{\doiref{10.1103/PhysRevD.80.104019}{Phys.~Rev.~D80,~104019~(2009)}},
\texttt{\arxivref{0905.2675}}.
%
\bibitem{Myers:2010jv}
R.~C.~Myers, M.~F.~Paulos and A.~Sinha,
\textit{``{Holographic studies of quasi-topological gravity}''},
\texttt{\arxivref{1004.2055}}.
%
\bibitem{Kulaxizi:2010jt}
M.~Kulaxizi and A.~Parnachev,
\textit{``{Energy Flux Positivity and Unitarity in CFTs}''},
\texttt{\arxivref{1007.0553}}.
%
\bibitem{Bousso:1999xy}
R.~Bousso,
\textit{``{A Covariant Entropy Conjecture}''},
\textsf{JHEP~9907,~004~(1999)},
\texttt{\arxivref{hep-th/9905177}}.
%
\bibitem{Bousso:2002ju}
R.~Bousso,
\textit{``{The holographic principle}''},
\textsf{\doiref{10.1103/RevModPhys.74.825}{Rev.~Mod.~Phys.~74,~825~(2002)}},
\texttt{\arxivref{hep-th/0203101}}.
%
\bibitem{Danielsson:2009gi}
U.~H.~Danielsson and L.~Thorlacius,
\textit{``{Black holes in asymptotically Lifshitz spacetime}''},
\textsf{\doiref{10.1088/1126-6708/2009/03/070}{JHEP~0903,~070~(2009)}},
\texttt{\arxivref{0812.5088}}.
%
\bibitem{Brynjolfsson:2010mk}
E.~J.~Brynjolfsson, U.~H.~Danielsson, L.~Thorlacius and T.~Zingg,
\textit{``{Holographic models with anisotropic scaling}''},
\texttt{\arxivref{1004.5566}}.
%
\bibitem{Bertoldi:2009vn}
G.~Bertoldi, B.~A.~Burrington and A.~Peet,
\textit{``{Black Holes in asymptotically Lifshitz spacetimes with arbitrary
  critical exponent}''},
\textsf{\doiref{10.1103/PhysRevD.80.126003}{Phys.~Rev.~D80,~126003~(2009)}},
\texttt{\arxivref{0905.3183}}.
%
\bibitem{Donos:2010ax}
A.~Donos, J.~P.~Gauntlett, N.~Kim and O.~Varela,
\textit{``{Wrapped M5-branes, consistent truncations and AdS/CMT}''},
\texttt{\arxivref{arXiv:1009.3805}},
* Temporary entry *.
%
\bibitem{Si:2001kx}
Q.~Si, S.~Rabello, K.~Ingersent and J.~L.~Smith,
\textit{``Locally critical quantum phase transitions in strongly correlated
  metals''},
\textsf{Nature~413,~804~(2001)},
\href{http://dx.doi.org/10.1038/35101507}{\texttt{http://dx.doi.org/10.1038/35%
101507}}.
%
\end{thebibliography}
\bibliographystyle{nb}

\end{document}